\documentclass[letterpaper, 10 pt, conference]{ieeeconf}  

\IEEEoverridecommandlockouts                              

\usepackage{amsmath} 
\usepackage{amssymb}  
\usepackage{bm}

\usepackage[latin1]{inputenc}
\usepackage{tikz}
\usetikzlibrary{shapes,arrows}
\usepackage{booktabs}
\usepackage{multirow}

\usepackage{comment}

\usepackage{colortbl} 
\usepackage{xcolor} 
\usepackage{soul}

\title{\LARGE \bf
Joint Parameterization of Hybrid Physics-Based and Machine Learning Li-Ion Battery Model
}

\author{Jackson Fogelquist and Xinfan Lin
\thanks{Jackson Fogelquist and Xinfan Lin are with the Department of Mechanical and Aerospace Engineering, University of California, Davis, CA 95616, USA, corresponding author e-mail: {\tt\small lxflin@ucdavis.edu}}%
}

\begin{document}

\maketitle
\thispagestyle{empty}
\pagestyle{empty}

\begin{abstract}

Electrochemical hybrid battery models have major potential to enable advanced physics-based control, diagnostic, and prognostic features for next-generation lithium-ion battery management systems.
This is due to the physical significance of the electrochemical model, which is complemented by a machine learning model that compensates for output prediction errors caused by system uncertainties.
While hybrid models have demonstrated robust output prediction performance under large system uncertainties, they are highly susceptible to the influence of uncertainties during parameter identification, which can compromise the physical significance of the model.
To address this challenge, we present a parameter estimation framework that explicitly considers system uncertainties through a discrepancy function.
The approach also incorporates a downsampling procedure to address the computational barriers associated with large time-series data sets, as are typical in the battery domain.
The framework was validated in simulation, yielding several mean parameter estimation errors that were one order of magnitude smaller than those of the conventional least squares approach.
While developed for the high-uncertainty, electrochemical hybrid modeling context, the estimation framework is applicable to all models and is presented in a generalized form.

\end{abstract}
\section{Introduction}

As lithium-ion (Li-ion) batteries become increasingly integrated into our society, it is critical to ensure the safe and high performance operation of these devices through next-generation \emph{advanced} battery management systems (BMSs).
The distinguishing feature of an advanced BMS is the use of a physics-based electrochemical battery model, as opposed to the traditional implementation of an empirical equivalent circuit model \cite{chaturvedi_algorithms_2010}.
Electrochemical models, such as the popular Doyle-Fuller-Newman (DFN) model \cite{doyle_modeling_1993}, use first principles to explicitly represent the physical electrochemical mechanisms within a battery.
The primary benefit of electrochemical models is the physical significance of their parameters and internal states, which enable various BMS capabilities, such as degradation monitoring through the estimation of health-related states and parameters \cite{allam_-line_2021}, optimal control and fault detection through the availability of physical states \cite{couto_faster_2022}, and enhanced feedback estimation and control through accurate output voltage predictions \cite{padisala_online_2024}.
The fundamental disadvantages of electrochemical models are their high dimensionality and nonlinearity, which make them prone to system uncertainties \cite{fogelquist2023eTran}, identifiability issues \cite{berliner_nonlinear_2021}, and considerable computational expense that inhibits implementation in online BMS applications \cite{lin2019modeling}. 

Significant research efforts have been directed toward addressing the challenges posed by the complexity of electrochemical models.  
Typically, these efforts involve the use of reduced-order electrochemical models, such as the single particle model (SPM) \cite{ning_cycle_2004}, which are simplified versions of full-order models that reduce complexity by carefully omitting the minor dynamics.
One solution that has shown great promise is the development of electrochemical hybrid models, in which a reduced-order electrochemical model is augmented with a machine learning (ML) model that compensates for system uncertainties (e.g., unmodeled dynamics, noise/bias in fitting data) based on patterns extracted from observations \cite{aykol_perspectivecombining_2021}.
The use of hybrid \emph{residual} models has been particularly successful, in which the ML model predicts the output voltage error of the electrochemical model (i.e., the residual) and applies it as a correction.

The first electrochemical hybrid residual model was developed in simulation in \cite{park_hybrid_2017}, in which a recurrent neural network was implemented to predict the residual between a full-order DFN model and reduced-order SPM.
In \cite{tu2021integrating}, state information from the electrochemical model was incorporated as an input to the ML residual model, which was found to significantly improve the residual prediction accuracy in simulation.
This work was continued in \cite{tu_integrating_2023} with a theoretical justification for the use of state information as an ML input.
The first experimental validation of an electrochemical hybrid residual model was recently conducted in \cite{pozzato_accelerating_2024}, in which the performance of several ML algorithms were compared for the prediction of the voltage hysteresis residual of a lithium-iron-phosphate cell.
Recently, we developed a hybrid modeling framework that utilizes a Gaussian process regression (GPR) residual model and training-data sampling procedure to improve computational efficiency, which was experimentally validated in \cite{fogelquist2024lightweight}.
These works demonstrate that the hybrid modeling approach is advantageous because the reduced-order electrochemical model substantially reduces computational expense while the ML model adequately compensates for system uncertainties (including those introduced by the reduced-order model), such that accurate output predictions may be delivered at an ample rate for BMS applications.

While electrochemical hybrid models may address the difficulties related to computational expense and system uncertainties, the parameter estimation challenge still remains to be solved---in fact, it becomes even more complex through the increased dimensionality that comes from introducing the ML model (hyper)parameters.
In addition, the hybrid framework introduces a new identifiability issue in that the output voltage may not indicate the accuracy of the parameter set, i.e., the voltage output of an electrochemical model with an erroneous parameterization may be appropriately compensated by the ML residual model such that it gives no indication of an inaccurate parameterization.
These challenges were not considered in the aforementioned works because the studies were either performed in simulation (where the true parameter values were known) \cite{park_hybrid_2017,tu2021integrating,tu_integrating_2023} or under an experimental parameterization subject to system uncertainties \cite{pozzato_accelerating_2024,fogelquist2024lightweight}, which have been shown to critically impact estimation accuracy \cite{fogelquist2023eTran,fogelquist_error_2023}.

The objective of this work is to develop a parameter estimation approach for electrochemical hybrid residual models that explicitly considers system uncertainties.
Our methodology is inspired by the Kennedy and O'Hagan calibration approach of incorporating system uncertainties in the estimation procedure by representing them as a Gaussian process \cite{kennedy_bayesian_2001}.
This is a significant contribution because it is the first attempt, to the best of our knowledge, to identify parameters of an electrochemical hybrid residual model while accounting for the influence of uncertainties.
While the parameter estimation framework was developed for hybrid residual models, we present it in a generalized form that is applicable to conventional models as well.
One important feature of the framework is a data sampling procedure that maintains the tractability of the algorithm, which is essential for battery applications where time-series data sets are often large.
During simulation validation, the proposed framework yielded mean electrochemical parameter estimation errors that were one order of magnitude smaller than those of the conventional least squares approach, indicating the importance and effectiveness of considering system uncertainties in the estimation procedure.
Thus, the proposed framework may facilitate the accurate parameterization of hybrid residual models to maintain physically-significant predictions of internal states and output voltage---the core data required for advanced BMS capabilities.

\section{Electrochemical Hybrid Residual Model}
\label{sec:hybrid_model}

This section provides a brief overview of the electrochemical hybrid residual modeling paradigm that the parameter estimation framework was developed for.
Electrochemical battery models, residual models, and the complete hybrid architecture are summarized as follows, with reference to the model developed in \cite{fogelquist2024lightweight} that will be implemented to validate the parameter estimation framework.

\subsection{Electrochemical Battery Model}

The prevailing electrochemical battery model is the DFN model \cite{doyle_modeling_1993}, which characterizes electrochemical phenomena using first principles, namely, mass conservation via Fick's second law of diffusion, charge conservation via Ohm's law, and charge transfer kinetics via the Butler-Volmer equation.
The resulting set of coupled partial differential equations must be implicitly solved through iterative methods, which renders considerable computational expense.
As a result, reduced-order electrochemical models have been developed to strike a balance between computational efficiency and model fidelity.
For our hybrid model, we selected the widely-adopted single particle model with electrolyte dynamics (SPMe) \cite{LaiJPS2020}, which was derived from the full-order DFN model through the simplification that reaction current density is uniform across each electrode.
As a result, the solid-phase lithium concentrations are uniform, allowing the electrode electrochemical mechanisms (e.g., diffusion, (de)intercalation) to be represented with a single particle.
The anode and cathode particles are then related through the electrolyte diffusion dynamics across the cell.

The SPMe predicts the battery internal states and output terminal voltage $V$ from the input current $I$.
The terminal voltage is expressed as
\begin{equation}
\label{eqn:SPMe_Vout}
\begin{split}
    V & = U_p(c_{se,p})-U_n(c_{se,n}) + \phi_{e,p}(c_{e,p}) - \phi_{e,n}(c_{e,n}) \\
    & + \eta_p(c_{se,p},c_{e,p}) - \eta_n(c_{se,n},c_{e,n}) - IR_{l},
\end{split}
\end{equation}
which captures the differences between the cathode and anode open-circuit potentials $U$, electrolyte potentials $\phi_e$, and overpotentials $\eta$, denoted by subscripts $p$ and $n$, respectively.
The final term captures the Ohmic voltage drop in function of the lumped resistance term $R_l$, which characterizes the net Ohmic resistance of the electrolyte, current collectors, and solid-electrolyte interphase layer.
The three potentials $(U,\phi_e,\eta)$ are nonlinear functions of the lithium concentration at the electrode particle surface $c_{se}$ and electrolyte boundary $c_e$, which are governed by the ionic diffusion dynamics through Fick's second law.
The electrode surface concentrations may be used to compute the surface state of charge (SOC), $SOC_{surf}(c_{se,p},c_{se,n})$, which is an important (lumped) state that conveys the fraction of available charge in the cell.
In addition, the electrode bulk lithium concentration states $\overline{c}_s$ are also commonly included in the SPMe formulation, which represent the total molar concentration of lithium in the electrode particles.
While $\overline{c}_s$ is not required to compute the terminal voltage, it allows an alternative definition of the SOC to be expressed, namely, the bulk SOC, $SOC_{bulk}(\overline{c}_{s,p},\overline{c}_{s,n})$.
The bulk SOC serves as a perfect integrator of the reaction current density, providing an additional perspective on battery performance.
The complete formulation of the SPMe is detailed in \cite{LaiJPS2020}.

\subsection{Residual Model}
\label{sec:HM_residual}

The purpose of the residual model is to predict the output residual for a given system input, based on a set of observed training data points.
Conventionally, neural networks have demonstrated excellent suitability for this task, though for our hybrid model in \cite{fogelquist2024lightweight}, we developed a computationally-efficient GPR residual model that is implemented in this work.
The following is a brief review of GPR models, based on \cite{rasmussen_gaussian_2006}.
The goal of a GPR model is to express an unknown continuous relationship between a series of observed system inputs $\bm{X}=[\bm{x}_1,\ldots,\bm{x}_N]$ and outputs $\bm{y}=[y_1,\ldots,y_N]$, where the inputs may be multivariate such that $\bm{x}_k=[x_{k,1},\ldots,x_{k,d}]^T$.  
Subscript $N$ denotes the number of observations and subscript $d$ denotes the number of input signals per observation $k$.
By considering the unknown input-output relationship to be a zero-mean Gaussian process, i.e., a collection of jointly Gaussian random variables $f(\bm{x})$ with a mean value of zero (the ideal value of a residual), 
\begin{equation}
    \label{eqn:GP}
    f(\bm{x}) \sim \mathcal{GP}(0,k(\bm{x},\bm{x}')),
\end{equation}
the relationship is entirely defined by the covariance function $k(\bm{x},\bm{x}')$.
We selected the versatile squared exponential covariance function for our application,
\begin{equation}
    \label{eqn:cov}
    k(\bm{x},\bm{x}') = \sigma^2_f\ \mathrm{exp}\left(-\frac{1}{2}(\bm{x}-\bm{x}')^T \bm{L} (\bm{x}-\bm{x}')\right),
\end{equation}
which is characterized by the scale factor hyperparameter $\sigma^2_f$ and matrix $\bm{L} = \mathrm{diag}(\bm{l})^{-2}$, where $\bm{l}=[l_1,\ldots,l_d]$ are the length scale hyperparameters associated with each input signal.
The value of $k(\bm{x},\bm{x}')$ ranges from zero to $\sigma^2_f$ and indicates the correlation between the input vectors $\bm{x}$ and $\bm{x}'$, i.e., $k(\bm{x},\bm{x}')$ is zero when the inputs are completely uncorrelated $(\bm{x}-\bm{x}'=\pm\bm{\infty})$ and $\sigma^2_f$ when they are perfectly correlated $(\bm{x}-\bm{x}'=\bm{0})$.
Elementwise evaluation of Eqn.\ (\ref{eqn:cov}) yields the covariance matrices associated with the input training data $\bm{X}$ and input test points $\bm{X}_*$, where $\bm{K} = \bm{K}(\bm{X},\bm{X})$, $\bm{K}_* = \bm{K}(\bm{X},\bm{X}_*)$, and $\bm{K}_{**} = \bm{K}(\bm{X}_*,\bm{X}_*)$.
To account for i.i.d. Gaussian measurement noise in the observations, the observation noise variance $\sigma^2_n$ is introduced to form the noisy training data covariance matrix $\bm{K}_n = \bm{K} + \sigma^2_n \bm{I}$. 

The joint distribution of the observed outputs $\bm{y}$ and function predictions $\bm{f}_*$ (associated with input test points $\bm{X}_*$) can be conditioned on the observed training data $(\bm{X},\bm{y})$ to yield the joint posterior distribution
\begin{equation}
    \bm{f}_* | \bm{X},\bm{y},\bm{X}_* \sim \mathcal{N}\left(\overline{\bm{f}}_*,\bm{cov}(\bm{f}_*)\right),
\end{equation}
where
\begin{align}
    \label{eqn:GPR_mean} 
    \overline{\bm{f}}_* &= \bm{K}_*^T\bm{K}_n^{-1} \bm{y}, \\
    \label{eqn:GPR_cov}
    \bm{cov}(\bm{f}_*) &= \bm{K}_{**} - \bm{K}_*^T\bm{K}_n^{-1} \bm{K}_* + \sigma^2_n \bm{I}.
\end{align}
Eqns.\ (\ref{eqn:GPR_mean}) \& (\ref{eqn:GPR_cov}) are the GPR prediction equations, in which a specified set of input test points $\bm{X}_*$ and training data $(\bm{X},\bm{y})$ yield predictions of the conditional mean outputs $\overline{\bm{f}}_*$ and covariance matrix $\bm{cov}(\bm{f}_*)$.
For more information about GPR, \cite{rasmussen_gaussian_2006} is recommended as an excellent source.

\subsection{Hybrid Model Architecture}
\label{sec:hybrid_model_arch}

\begin{figure} [!ht]
    \centering
    \includegraphics[width=2.6in]{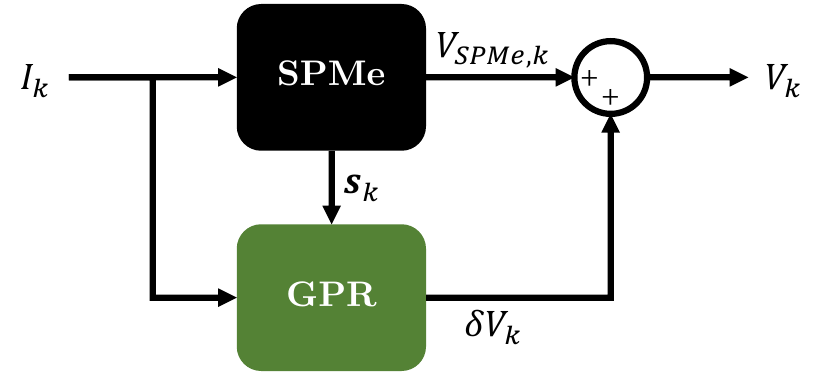}
    \caption{Electrochemical hybrid residual model schematic \cite{fogelquist2024lightweight}.}
    \label{fig:HM_schem} 
\end{figure}

The purpose of the hybrid model is to encode the physical mechanisms of battery operation (e.g., lithium diffusion, reaction kinetics) via the electrochemical model while compensating for prediction errors through the ML residual model, based on past observations.
At a given time step, the input current is applied to the electrochemical and residual models, from which the electrochemical model predicts the internal physical states and output voltage.
A subset of the internal states is provided as an input to the residual model, which yields a prediction of the voltage residual.
The residual and electrochemical voltage predictions are then summed to deliver the final output voltage.
A schematic of the SPMe-GPR hybrid model from \cite{fogelquist2024lightweight} is presented in Fig.\ \ref{fig:HM_schem}, in which the input current $I$, SPMe voltage prediction $V_{SPMe}$, subset of internal states $\bm{s}$, residual prediction $\delta V$, and final output voltage $V$ are labeled at time step $k$.
In this implementation, $\bm{s}$ comprises the surface SOC, bulk SOC, and anode electrolyte concentration, which were selected to succinctly encode the information contained in the six SPMe concentration states, i.e., $\bm{s} = [SOC_{surf}(c_{se,p},c_{se,n}),SOC_{bulk}(\overline{c}_{s,p},\overline{c}_{s,n}),c_{e,n}]^T$ (note that $c_{e,n}$ is nearly symmetric to $c_{e,p}$, allowing a single electrolyte concentration state to represent the electrolyte gradient).
Therefore, the complete input vector to the GPR residual model is $\bm{x} = \left[\begin{smallmatrix}I\\\bm{s}\end{smallmatrix}\right] = [I,SOC_{surf},SOC_{bulk},c_{e,n}]^T$, which is applied in the covariance function in Eqn.\ (\ref{eqn:cov}).
The reader is referred to \cite{fogelquist2024lightweight} for more details.

\section{Parameter Estimation Framework}
\label{sec:est_framework}

The parameter estimation framework is an offline methodology that is inspired by the model calibration approach proposed by Kennedy and O'Hagan \cite{kennedy_bayesian_2001}, in which the discrepancies between simulation model predictions and physical measurements (i.e., the voltage residual) are represented as a Gaussian process and integrated into the estimation procedure.
A frequentist formulation of the original Bayesian formulation was presented in \cite{tuo_theoretical_2016}, in which the physical system and simulation model were assumed to be deterministic.
As battery systems and models are conventionally treated as deterministic, we adopted this formulation and express the measured physical system output $y^m$ as the sum of the model output $y$ and discrepancy $\delta$ at each time step $k$,
\begin{equation}
    \label{eqn:cal_model}
    y^m_k = y_k(\bm{u}_k,\bm{\theta})+\delta_k(\bm{x}_k(\bm{u}_k,\bm{\theta}),\bm{\gamma}).
\end{equation}
Here, $\bm{u}_k=[u_1,\ldots,u_k]^T$ is the input excitation sequence vector, $\bm{\theta}$ is the vector of model parameters, $\bm{x}_k=[x_{k,1},\ldots,x_{k,d}]^T$ is the vector of inputs to the discrepancy function (which may be internal states from the model and thus depend on $\bm{u}_k$ and $\bm{\theta}$), and $\bm{\gamma}$ is the vector of discrepancy function hyperparameters.
The physical output, model output, and discrepancy may also be represented as sequence vectors across the $N$ time steps, such that $\bm{y}^m=[y^m_1,\ldots,y^m_N]^T$, $\bm{y}=[y_1,\ldots,y_N]^T$, and $\bm{\delta}=[\delta_1,\ldots,\delta_N]^T$.

The discrepancy $\bm{\delta}$ is unknown and thus modeled as a zero-mean Gaussian process, according to Eqn.\ (\ref{eqn:GP}), which is characterized by the squared exponential covariance function in Eqn.\ (\ref{eqn:cov}), as in \cite{kennedy_bayesian_2001,tuo_theoretical_2016}, among others.
Note that other covariance functions may be used, but we selected the squared exponential due to its versatility in the literature and consistency with our hybrid model formulation.
Elementwise evaluation of the covariance function under a given set of input data $\bm{X}(\bm{u},\bm{\theta})$ and hyperparameters $\bm{\gamma}$ yields the covariance matrix $\bm{K}$.
We assume that the system output will contain i.i.d. Gaussian measurement noise and incorporate the observation variance $\sigma^2_n$ to form the noisy covariance matrix $\bm{K}_n = \bm{K} + \sigma^2_n \bm{I}$ (as in Section \ref{sec:HM_residual}).
By letting $\bm{K}=\sigma^2_f \bm{\Phi}$, the covariance scale factor $\sigma^2_f$ may be factored out of $\bm{K}_n$ such that
\begin{equation}
    \bm{K}_n = \bm{K} + \sigma^2_n \bm{I} = \sigma^2_f (\bm{\Phi}+\tilde{\sigma}^2_n \bm{I})=\sigma^2_f \bm{\Phi}_n,
\end{equation}
where $\bm{\Phi}_n=\bm{\Phi}+\tilde{\sigma}^2_n \bm{I}$ is the unscaled noisy covariance matrix and $\tilde{\sigma}^2_n=\sigma^2_n/\sigma^2_f$ is a dimensionless hyperparameter.
Thus, the discrepancy is represented as
\begin{equation}
    \bm{\delta}(\sigma^2_f,\bm{u},\bm{\theta},\bm{\gamma}) \sim \mathcal{N}\left(\bm{0},\sigma^2_f \bm{\Phi}_n(\bm{u},\bm{\theta},\bm{\gamma})\right),
\end{equation}
where the covariance matrix hyperparameters are $\bm{\gamma}=[\tilde{\sigma}^2_n,l_1,\ldots,l_d]$.

In addition, we introduce the model error $\epsilon$ between the output measurements and model predictions
\begin{equation}
    \epsilon_k(y^m_k,\bm{u}_k,\hat{\bm{\theta}}) = y^m_k - y_k(\bm{u}_k,\hat{\bm{\theta}}),
\end{equation}
which can be computed for a given set of input-output measurements $(\bm{u},\bm{y}^m)$ and parameter set $\hat{\bm{\theta}}$.
The error sequence vector follows as $\bm{\epsilon}=[\epsilon_1,\ldots,\epsilon_N]^T$.
As in \cite{tuo_theoretical_2016}, we employ maximum likelihood estimation to determine the model parameter set $\hat{\bm{\theta}}$ and discrepancy function hyperparameter set $\hat{\bm{\gamma}}$ that maximize the log likelihood of predicting the model error $\bm{\epsilon}$ with the discrepancy function $\bm{\delta}$, where the log likelihood is expressed as
\begin{equation}
    \label{eqn:loglikelihood}
    \begin{split}
        \log\ &p(\bm{\epsilon} | \hat{\sigma}^2_f,\bm{y}^m,\bm{u},\hat{\bm{\theta}},\hat{\bm{\gamma}}) = \\
        &-\frac{N}{2} \log 2\pi\hat{\sigma}^2_f-\frac{1}{2}\log\left|\bm{\Phi}_n(\bm{u},\hat{\bm{\theta}},\hat{\bm{\gamma}})\right|\\
        &-\frac{1}{2\hat{\sigma}^2_f}\bm{\epsilon}(\bm{y}^m,\bm{u},\hat{\bm{\theta}})^T \bm{\Phi}_n^{-1}(\bm{u},\hat{\bm{\theta}},\hat{\bm{\gamma}}) \bm{\epsilon}(\bm{y}^m,\bm{u},\hat{\bm{\theta}}).
    \end{split}  
\end{equation}

Taking the derivative of Eqn.\ (\ref{eqn:loglikelihood}) with respect to $\hat{\sigma}^2_f$ and equating it to zero (i.e., applying the first-order optimality condition to the $\hat{\sigma}^2_f$ dimension) yields the analytical solution
\begin{equation}
    \label{eqn:sigma2f}
    \hat{\sigma}^2_f = \frac{1}{N}\bm{\epsilon}(\bm{y}^m,\bm{u},\hat{\bm{\theta}})^T \bm{\Phi}_n^{-1}(\bm{u},\hat{\bm{\theta}},\hat{\bm{\gamma}}) \bm{\epsilon}(\bm{y}^m,\bm{u},\hat{\bm{\theta}}).
\end{equation}
Inserting Eqn.\ (\ref{eqn:sigma2f}) into the log likelihood function in Eqn.\ (\ref{eqn:loglikelihood}) provides the estimation objective function, which is independent of $\hat{\sigma}^2_f$,
\begin{equation}
    \label{eqn:objective}
    \begin{split}
        \max_{\hat{\bm{\theta}},\hat{\bm{\gamma}}}\ &\log p(\bm{\epsilon} | \bm{y}^m,\bm{u},\hat{\bm{\theta}},\hat{\bm{\gamma}}) = \\ 
        &-\frac{N}{2} \log \frac{2\pi}{N}\bm{\epsilon}(\bm{y}^m,\bm{u},\hat{\bm{\theta}})^T \bm{\Phi}_n^{-1}(\bm{u},\hat{\bm{\theta}},\hat{\bm{\gamma}}) \bm{\epsilon}(\bm{y}^m,\bm{u},\hat{\bm{\theta}})\\
        &-\frac{1}{2}\log\left|\bm{\Phi}_n(\bm{u},\hat{\bm{\theta}},\hat{\bm{\gamma}})\right|-\frac{N}{2}.
    \end{split}  
\end{equation}

Interestingly, Eqn.\ (\ref{eqn:objective}) can be rewritten as
\begin{equation}
    \label{eqn:objective_simp}
    \begin{split}
        &\min_{\hat{\bm{\theta}},\hat{\bm{\gamma}}}\ J =\\
        &\left|\bm{\Phi}_n(\bm{u},\hat{\bm{\theta}},\hat{\bm{\gamma}})\right|^{\frac{1}{N}}\bm{\epsilon}(\bm{y}^m,\bm{u},\hat{\bm{\theta}})^T \bm{\Phi}_n^{-1}(\bm{u},\hat{\bm{\theta}},\hat{\bm{\gamma}}) \bm{\epsilon}(\bm{y}^m,\bm{u},\hat{\bm{\theta}}),
    \end{split}
\end{equation}
which is similar in form to the conventional least squares objective
\begin{equation}
    \label{eqn:objective_LS}
        \min_{\hat{\bm{\theta}}}\ J =\bm{\epsilon}(\bm{y}^m,\bm{u},\hat{\bm{\theta}})^T \bm{\epsilon}(\bm{y}^m,\bm{u},\hat{\bm{\theta}}).
\end{equation}
The objective function in Eqn.\ (\ref{eqn:objective_simp}) utilizes the inverse of the unscaled covariance matrix $(\bm{\Phi}_n^{-1})$ to weight the elements of the model error inner product.
In addition, the objective is scaled by the determinant of the covariance matrix, which yields a smaller value when the off-diagonal terms are closer to unity, as associated with more strongly correlated input vectors $(\bm{x},\bm{x}')$, according to Eqn.\ (\ref{eqn:cov}).
In fact, considering the converse case of completely uncorrelated input vectors (i.e., $\bm{\Phi}_n=a\bm{I}$) reveals that the objective function in Eqn.\ (\ref{eqn:objective_simp}) becomes the least squares objective in Eqn.\ (\ref{eqn:objective_LS}).
Thus, the new objective function favors (hyper)parameter estimates that increase the correlation of the input signals while reducing the errors between the model predictions and measurements.

A limitation of the new objective function is that the computational expense of inverting $\bm{\Phi}_n$ scales cubically with its dimension \cite{rasmussen_gaussian_2006}, i.e., the number of data points in $\bm{\epsilon}$ that are used to fit the model to the measurements.
This quickly becomes intractable in battery applications, as voltage time-series data sets often contain thousands of observations.
To address this, we implement a data sampling procedure, originally proposed in \cite{fogelquist2024lightweight}, through which $\bm{\epsilon}$ is downsampled by utilizing a specified number of evenly distributed points in the sequence.
The input vectors at the associated time steps are then used to construct $\bm{\Phi}_n$.
In this work, we select 300 samples from $\bm{\epsilon}$ to yield a computationally efficient 300 $\times$ 300 element covariance matrix.

An important consideration is that the objective function in Eqn.\ (\ref{eqn:objective}) may not have a unique solution, as multiple combinations of model parameters and discrepancy function hyperparameters may fit the data equally well \cite{kennedy_bayesian_2001,tuo_theoretical_2016}.
While this is an established trait of log-likelihood maximization for GPR hyperparameter tuning \cite{rasmussen_gaussian_2006}, it can be plainly seen by examining the original model in Eqn.\ (\ref{eqn:cal_model}), as it is possible for different values of $\bm{\theta}$ and $\bm{\gamma}$ to yield model and discrepancy function outputs that sum to match the measurements.
Nevertheless, the parameter estimates under this objective tend to be accurate, as demonstrated in Section \ref{sec:sim_val}.

Finally, we emphasize that the objective function in Eqn.\ (\ref{eqn:objective}) is not limited to the hybrid modeling context, but is applicable to all parameter estimation problems.
The discrepancy function simply represents the misalignment between the model predictions and reality, and implementation only requires the construction of the covariance matrix $\bm{\Phi}_n$ from a set of input signals.
However, if a hybrid model is used and configured with a residual model that matches the form of the discrepancy function, as in the hybrid model presented in Section \ref{sec:hybrid_model}, then the framework has the additional advantage of simultaneously tuning the residual model hyperparameters during the parameter estimation, which can simplify the training procedure.

\section{Simulation Validation}
\label{sec:sim_val}

To validate the parameter estimation framework, we set up a simulation study in which six health-related electrochemical SPMe parameters were estimated from simulated ``measurement" data generated from the full-order DFN model.
The true parameter values are thus known from their implementation in the DFN model and were used for evaluating the performance of the estimation framework.
The model uncertainty is characterized by the mismatch in voltage outputs due to the unmodeled dynamics in the reduced-order SPMe.
Measurement uncertainty was also incorporated, in which i.i.d. Gaussian measurement noise with a mean and standard deviation of 10 mV was added to the DFN voltage outputs.
The DFN model and SPMe were configured with a parameter set adapted from \cite{li_data-driven_2022}. 
However, in the estimation scenario, the true values of the six target parameters are unknown.
To incorporate this uncertainty, initial errors in the range of $\pm(50\%-100\%)$ were randomly generated and applied to each target parameter in the SPMe.

The target parameters for estimation were the anode and cathode active material volume fractions, $\varepsilon_{s,n}$ and $\varepsilon_{s,p}$, anode and cathode solid-phase diffusion coefficients, $D_{s,n}$ and $D_{s,p}$, electrolyte diffusion coefficient $D_e$, and electrolyte volume fraction $\varepsilon_e$.
Physically, $\varepsilon_{s,n}$ and $\varepsilon_{s,p}$ represent the proportion of each electrode volume that is capable of storing lithium (and thus participating in the electrochemical reaction), while $\varepsilon_e$ indicates the average fraction of the electrode volume that is occupied by electrolyte.
The diffusion coefficients $D_{s,n}$, $D_{s,p}$, and $D_e$ characterize the rate at which lithium diffuses through the anode solid particles, cathode solid particles, and electrolyte, respectively.
Several studies have shown that these six parameters decrease throughout the life of a battery \cite{birkl_degradation_2017,zhou_impedance_2019}, as their values are linked to various electrochemical degradation mechanisms.
Thus, these parameters were selected for estimation because a BMS would need to iteratively estimate them throughout the life of a battery to maintain prediction accuracy and monitor the state of health.

The target parameters were organized into three groups of two, according to their relative output voltage sensitivities.
The groups are listed in Table \ref{tab:estResults}, where Group 1 contains the most sensitive parameters, followed by Group 2, and then Group 3, based on the sensitivity rankings in \cite{allam_-line_2021,li_data-driven_2022}.
The parameters in each group were jointly estimated, beginning with the most sensitive group and ending with the least.
This is typical practice for estimation problems with a significant number of target parameters, as jointly estimating all of the parameters together may make the problem ill-posed (i.e., the data do not contain sufficient information to identify all of the parameters) \cite{fogelquist2023eTran,berliner_nonlinear_2021,fogelquist_error_2023} and cause large estimation errors in the weakly sensitive parameters due to relatively small errors in the strongly sensitive parameters \cite{li_data-driven_2022}.

Table \ref{tab:estResults} also indicates the input current profiles that were used for each estimation.
Since the parameters in Group 1 are related to the open circuit voltage, a slow (0.5C) constant-current discharge profile was selected to generate voltage data.
The Group 2 parameters are related to solid-phase diffusion, which occurs on a relatively long time scale; thus, a fast (5C) constant-current discharge profile was selected to establish and maintain large particle concentration gradients.
As Group 3 contains the electrolyte parameters, a square wave (pulse) profile was selected to provide continuous excitation, as the electrolyte dynamics typically operate on a relatively short time scale.
The sequential estimation of the three groups was repeated for three iterations to allow the estimates to converge.

The objective function in Eqn.\ (\ref{eqn:objective}) was implemented for the estimation, in which the input vectors for the covariance matrix $\bm{\Phi}_n$ contained the input current, surface SOC, bulk SOC, and anode electrolyte concentration, i.e., $\bm{x} =  [I,SOC_{surf},SOC_{bulk},c_{e,n}]^T$, based on the configuration of the GPR residual model in Section \ref{sec:hybrid_model}.
Particle swarm optimization was employed to maximize the objective function because of its non-concavity.
In addition, the performance of the framework was found to be weakly sensitive to the dimensionless hyperparameter $\tilde{\sigma}^2_n$, so it was fixed at a small value of 0.1 to reduce the number of design variables.
The estimation was repeated under the conventional least squares objective in Eqn.\ (\ref{eqn:objective_LS}), and the results are listed in Table \ref{tab:estResults}.

\begin{table*}
    \caption{Estimation Errors under New Approach \& Least Squares}
    \renewcommand{\arraystretch}{1.3}
    \label{tab:estResults}
    \centering
    \begin{tabular}{cccccc}
        \arrayrulecolor{black} \hline
         &  &  &  & \multicolumn{2}{c}{\underline{Estimation Error}} \\        
        Group & Estimation Profile & Parameter & Initial Error & New Approach & Least Squares \\
        \hline
        \multirow{2}{*}{1} & \multirow{2}{*}{0.5C Discharge} & $\varepsilon_{s,n}$ & -95.9\% & -0.01\% & -0.01\% \\
         &  & $\varepsilon_{s,p}$ & 93.5\% & 0.75\% & 2.02\% \\
         \arrayrulecolor[gray]{.7} \hline
        \multirow{2}{*}{2} & \multirow{2}{*}{5C Discharge} & $D_{s,n}$ & -75.9\% & -10.32\% & -19.66\% \\
         &  & $D_{s,p}$ & -80.6\% & -5.62\% & 18.82\% \\
         \hline
        \multirow{2}{*}{3} & \multirow{2}{*}{1C Pulse (1/60 Hz)} & $D_{e}$ & 86.9\% & -7.27\% & 19.90\% \\
         &  & $\varepsilon_{e}$ & 59.4\% & 32.37\% & 29.90\% \\
         \arrayrulecolor{black} \hline
    \end{tabular}
\end{table*}

Table \ref{tab:estResults} reveals that the errors under the new approach are generally low $(\leq 10.3\%)$ with the exception of $\varepsilon_e$ $(32.4\%)$, which is reasonable given its very weak sensitivity.
The new approach yielded equivalent or higher estimation accuracy than least squares for all parameters except $\varepsilon_e$, which had a similar error.
The estimation errors from both approaches were much smaller than the initial errors (i.e., the randomly-assigned uncertainties), indicating convergence toward the true parameter values.

\begin{figure} [!ht]
    \centering
    \includegraphics[width=3.3in]{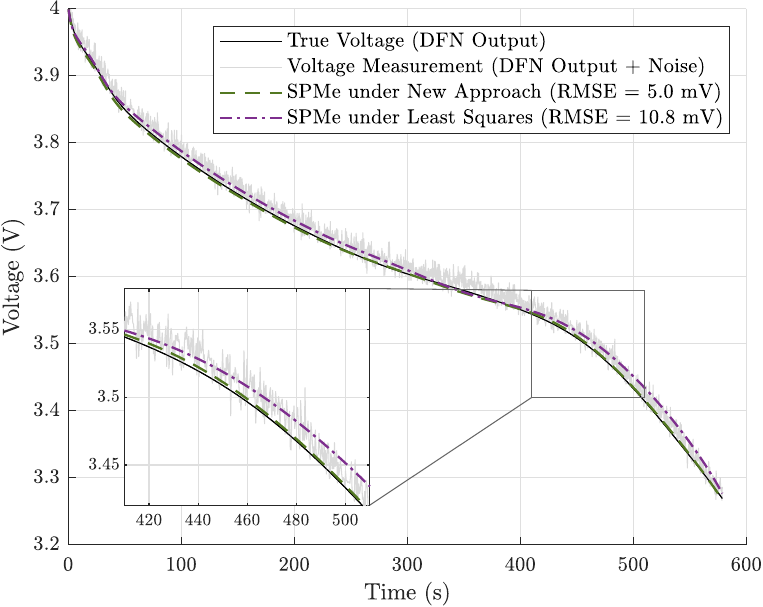} 
    \caption{SPMe voltage predictions under 5C discharge current profile.}
    \label{fig:V_comparison} 
\end{figure}

To visualize the performance of the model, Fig.\ \ref{fig:V_comparison} shows the SPMe voltage predictions for the 5C discharge current profile under the estimated parameters from the new approach (green dashed line) and least squares approach (purple dash-dotted line).
The voltage predictions are plotted alongside the true system voltage (i.e., the DFN model output under the true parameter values, shown in black) and the voltage measurement data (i.e., the DFN model output with injected measurement noise, shown in gray).
The root-mean-square errors (RMSEs) between the SPMe predictions and true voltage values are 5.0 mV under the new approach and 10.8 mV under the least squares approach.
This significant reduction in prediction error reveals the major advantage of the new approach---the ingrained discrepancy function correctly separates the true voltage from the voltage residual caused by uncertainties.
Thus, the new approach enables a parameterization that is fitted to the true system output rather than an output distorted by noise and other uncertainties, which is essential for maintaining the physical significance of the model (e.g., for degradation monitoring and internal state prediction).
On the other hand, the least squares approach fits the parameters to the voltage measurements (as opposed to the true voltage values), which is not necessarily desirable given the often-unpredictable behavior of the underlying uncertainties (e.g., sensor noise is random and can change over time, thus skewing the parameter estimates).
However, since noisy measurement data is typically used for feedback state estimation and control, the deviations between the measurements and true system outputs must still be considered.
The capability for the new approach to separate the true voltage from the voltage residual enables the training of an accurate residual model to compensate for system uncertainties, as discussed in Section \ref{sec:hybrid_model_arch} and demonstrated in \cite{fogelquist2024lightweight}.
Finally, it is noteworthy that the new approach uses the same information as least squares to perform the estimation, but is able to consider the influence of system uncertainties through the incorporation of the (unknown) discrepancy function.

The study was repeated eight times under different randomly-generated initial errors in the target parameters and injected noise profiles (still drawn from a Gaussian distribution with 10 mV mean and standard deviation).
The resulting estimation error means and standard deviations are presented in Table \ref{tab:estResults_stats} for each parameter and objective function.
The new approach generally delivered smaller mean errors than least squares, with nearly equivalent values for $\varepsilon_{s,n}$ and $\varepsilon_e$.
The largest improvements were observed in $\varepsilon_{s,p}$ and $D_{s,n}$, which have mean errors that are one order of magnitude smaller. 
The standard deviations of the errors are larger under the new approach, especially for the electrolyte parameters in Group 3.
The error variance appears to be inversely correlated with the output sensitivity of the parameters, as it increases as parameter sensitivity decreases, which is consistent with the definition of the Cram\'er-Rao bound.
The consequences of large error fluctuations in weakly sensitive parameters are minimal when predicting the output voltage, as the output is only subtly influenced by these parameters.
However, error fluctuations may present a challenge in health-monitoring applications when the trends in these parameters must be accurately tracked over time.
Still, since the mean errors are generally small, the large variances might be remedied by filtering the estimated parameter trends, e.g., with a moving average.
Regardless, we envision that the error variance could be reduced through data design, by optimizing estimation profiles to maximize parameter sensitivity---a topic for future work.
Overall, the performance of the new approach demonstrates that estimation accuracy can be improved by explicitly incorporating system uncertainties in the estimation procedure. 
The framework is thus validated as a viable method for addressing the identification challenges associated with electrochemical hybrid models.

\begin{table*}
    \caption{Estimation Error Means \& Standard Deviations}
    \renewcommand{\arraystretch}{1.3}
    \label{tab:estResults_stats}
    \centering
    \begin{tabular}{cccccc}
        \arrayrulecolor{black} \hline
         &  & \multicolumn{2}{c}{\underline{Mean Estimation Error}} & \multicolumn{2}{c}{\underline{Std.\ Dev.\ of Estimation Error}} \\
        Group & Parameter & New Approach & Least Squares & New Approach & Least Squares \\
        \hline
        \multirow{2}{*}{1} & $\varepsilon_{s,n}$ & 0.05\% & -0.01\% & 0.10\% & 0.02\% \\
         & $\varepsilon_{s,p}$ & -0.08\% & 2.03\% & 0.39\% & 0.02\% \\
         \arrayrulecolor[gray]{.7} \hline
        \multirow{2}{*}{2} & $D_{s,n}$ & -1.64\% & -18.94\% & 9.87\% & 1.07\% \\
         & $D_{s,p}$ & 3.83\% & 15.60\% & 8.80\% & 4.77\% \\
         \hline
        \multirow{2}{*}{3} & $D_{e}$ & 8.71\% & 11.93\% & 60.89\% & 6.20\% \\
         & $\varepsilon_{e}$ & 33.22\% & 32.86\% & 11.18\% & 2.60\% \\
         \arrayrulecolor{black} \hline
    \end{tabular}
\end{table*}

\section{Conclusions}

In this work, a parameter estimation approach was presented for the identification of physics-based electrochemical models in the hybrid battery modeling context, though it remains generalized to conventional estimation problems as well.
The methodology is inspired by the Kennedy and O'Hagan calibration approach \cite{kennedy_bayesian_2001}, in which a Gaussian process discrepancy function is integrated into the estimation procedure to explicitly consider system uncertainties.
Our approach also features a downsampling procedure to overcome the tractability issues associated with large time-series data sets, which are common in the battery domain.

The estimation performance of the new approach was compared with that of the conventional least squares approach in a simulation study, which incorporated modeling, measurement, and parameter uncertainties.
The new approach generally improved the estimation accuracy, yielding several mean estimation errors that were one order of magnitude smaller than those of the least squares approach.
However, error variance was larger under the new approach, which was attributed to a stronger dependence on parameter output sensitivity than the conventional approach.
Finally, the voltage prediction accuracy of the battery model was compared under the parameterizations from each approach, and the new approach yielded a closer fit to the true system output with a 54\% smaller RMSE.
Thus, the presented framework was validated as an effective means for enhancing estimation accuracy in modeling environments subject to substantial system uncertainties, such as electrochemical battery models.

The estimation framework is expected have special relevance for hybrid residual models, which are designed to compensate for the impacts of system uncertainties in their output predictions, but currently do not consider these uncertainties in their identification.
We thus plan to apply this framework in an experimental hybrid modeling context for further validation, and intend to explore the design of input excitations to enhance performance.

\addtolength{\textheight}{-0cm}   




\section*{Acknowledgment}

We appreciate the funding support from the NSF CAREER Program (Grant No. 2046292) and the NASA HOME Space Technology Research Institute (Grant No. 80NSSC19K1052).


\bibliographystyle{IEEEtran_nourls.bst} 
\bibliography{refs}

\end{document}